\documentclass[journal = apchd5, manuscript = article]{achemso}
\usepackage[english]{babel}
\usepackage{graphicx}
\usepackage{amsmath,amssymb,bm}
\usepackage[utf8]{inputenc}
\usepackage{natbib}
\usepackage{dsfont}
\usepackage{hyperref}
\usepackage[dvipsnames]{xcolor}
\usepackage{soul}
\usepackage{url}
\hypersetup{colorlinks=true,citecolor={blue},linkcolor={blue},urlcolor={blue}}
\usepackage{balance}
\usepackage{textcomp}
\usepackage{array}
\DeclareMathOperator{\sgn}{sgn}
\usepackage{upgreek}
\renewcommand{\mu}{\upmu}
\newcommand{\ium}{\mu\textrm{m}^{-1}}

\author{H. Sigurdsson}
\email{helg@hi.is}
\affiliation{Science Institute, University of Iceland, Dunhagi-3, IS-107 Reykjavik, Iceland}
\author{O. Kyriienko}
\affiliation{NORDITA, KTH Royal Institute of Technology and Stockholm University, Roslagstullsbacken 23, SE-106 91 Stockholm, Sweden}
\author{K. Dini}
\affiliation{ITMO University, Kronverkskiy prospekt 49, Saint Petersburg 197101, Russia}
\author{T. C. H. Liew}
\affiliation{Division of Physics and Applied Physics, School of Physical and Mathematical Sciences, Nanyang Technological University, 21 Nanyang Link, Singapore 637371}

\title{All-to-all connected networks by multi-frequency excitation of polaritons}

\makeindex
\begin{document}
\date{\today}

\begin{abstract}
We analyze theoretically a network of all-to-all coupled polariton modes, realized by a trapped polariton condensate excited by a comb of different frequencies. In the low-density regime the system dynamically finds a state with maximal gain defined by the average intensities (weights) of the excitation beams, analogous to active mode locking in lasers, and thus solves a maximum eigenvalue problem set by the matrix of weights. The method opens the possibility to tailor a superposition of populated bosonic modes in the trapped condensate by appropriate choice of drive.
\end{abstract}

Control over bosonic light-matter systems known as exciton-polariton condensates~\cite{CarusottoCiuti,Byrnes2014}, has increased dramatically over the recent years making them an excellent condensed matter candidate to study open many-body systems within semiclassical mean field theory. Exciton-polariton (or simply {\it polariton}) condensates can be generated either via resonant (coherent) excitation using an optical beam which creates a coherent ensemble of polaritons at a given energy and momenta, or by nonresonant excitation. The latter initially creates a reservoir of excitonic states, which, at high enough intensities, can macroscopically populate a lower energy polariton state via bosonic stimulated scattering. Polariton condensates can then be typically described by a macroscopic wave function governed by the appropriate mean-field dynamical equations which account for gain, dissipation, reservoir blueshift, and interactions between polaritons.
Of much interest is the possible application of polaritons in optoelectronic devices~\cite{Liew2011, Sanvitto_NatMat2016, Fraser2017}, such as all-optical logic~\cite{Espinosa-Ortega2013}, switches~\cite{Amo_NatPho2010, Giorgi_PRL2012, Gao_PRB2012, Cerna_NatComm2013, Grosso_PRB2014, Dreismann_NatMat2016}, and lasers~\cite{Kasprzak_Nature2006, Christopoulos_PRL2007, Das_PRL2011, Li_PRL2013, Su_NanoLett2017}. Recently, polaritonic lattices have also drawn attention as analog simulators~\cite{Ohadi_PRL2017,Berloff_NatMat2017,Sigurdsson2017}, where the steady-state solution for the driven-dissipative polariton lattice emulates a classical system of interacting spins. 

Being analogous to optical networks~\cite{Yamamoto_NatPho2014,Yamamoto_NatPho2016,Inagaki_Science2016}, the prospect of using polariton condensates for analog computing~\cite{AnalogComputing_Ulmann2003}, which relies on solving continuous time dynamics rather than operating digitally through a universal set of logic gates, could lead to new hybrid analog-digital computation devices with both fast analog simulation and digital accuracy. As an example, research devoted to spatial graphs of coupled polariton condensates~\cite{Tosi_NatPhys2012, Cristofolini_PRL2013, Ohadi_PRX2016, Lagoudakis_NJP2017} has revealed their ability to interfere and phase lock through inter-modal interactions~\cite{Baas_PRL2008, Lagoudakis_PRL2010} in the process of finding an optimal state which minimizes decay. The coherent coupling strength between neighbouring condensates is then tunable by either changing the separation distance, potential barriers between them, or the excitation strength~\cite{Ohadi_PRL2017, Berloff_NatMat2017}. On the other hand, going beyond nearest neighbour type coupling is a non-trivial task.

In this paper we study theoretically a system of driven-dissipative polaritonic modes linearly coupled to each other with an all-to-all connectivity. As a possible realization for the setup we propose a trapped polariton condensate in a microcavity driven by overlapping tight nonresonant beams modulated at multiple discrete frequencies. The system can be described by coupled differential equations for the internal modes and is similar to active mode locking in laser systems~\cite{Haus_IEEEJSTQE2000}. We show that the emergent network dynamically optimizes linear problems imprinted by the intensities (weights) of the excitation beams. Namely, it can find the maximal eigenvalue of the matrix imprinted by the weights. Similarly to the application of continuous Hopfield networks in optimizing complex problems through the Lyapunov function~\cite{Aiyer_IEEE1990, Talavan_NeurNet2002}, the system optimizes the Lyapunov exponent (Lyapunov energy, or net gain) in order to condense. Physically, this optimization comes from the bosonic stimulated scattering where the rate of polaritons populating a state increases with its occupation number. By controlling the weights associated with the oscillating nonresonant excitation one can tailor the distribution of bosons in each polariton mode making up the condensate.

\section{Theory}
We work with a scalar order parameter $\Psi$ corresponding to a macroscopic coherent field of polaritons. For simplicity, we will consider a one-dimensional system such as a polariton microwire as studied by several groups \cite{Wertz_NatPhys2010, Gao_PRB2012, Duan_APL2013,Sich_PRL2018}. However, our results are readily generalized to higher dimensions. The one dimensional Hamiltonian $\hat{H}_0$ reads
\begin{equation}
\hat{H}_0 = -\frac{\hbar }{2m^*} \frac{\partial^2}{\partial x^2} + V(x),
\end{equation}
where $m^*$ denotes the effective mass of lower branch exciton-polaritons, $V(x)$ corresponds to the introduced confinement potential, and we choose to work in frequency units. The system is characterized by time independent eigenstates $\varphi_n(x)$ of $\hat{H}_0$, which in general have a non-degenerate and monotonically varying discrete spectrum, $\hbar\omega_n$. 

We now consider the nonlinear Schr\"odinger equation describing a condensate of polaritons nonresonantly driven by a superposition of time-dependent tightly spatially localized non-resonant pumps $P(x,t)$ [see Fig.~\ref{fig.scheme}]. Low energy polaritons are then fed into the system by scattering from an active excitonic reservoir~\cite{Wouters_PRL2007} induced by the pump. Assuming that the reservoir relaxes much faster than the polariton condensate we can apply a quasi-stationary approximation which allows us to write a single equation of motion for the polariton condensate:
\begin{figure}[t!]
\centering
\includegraphics[width=0.7\linewidth]{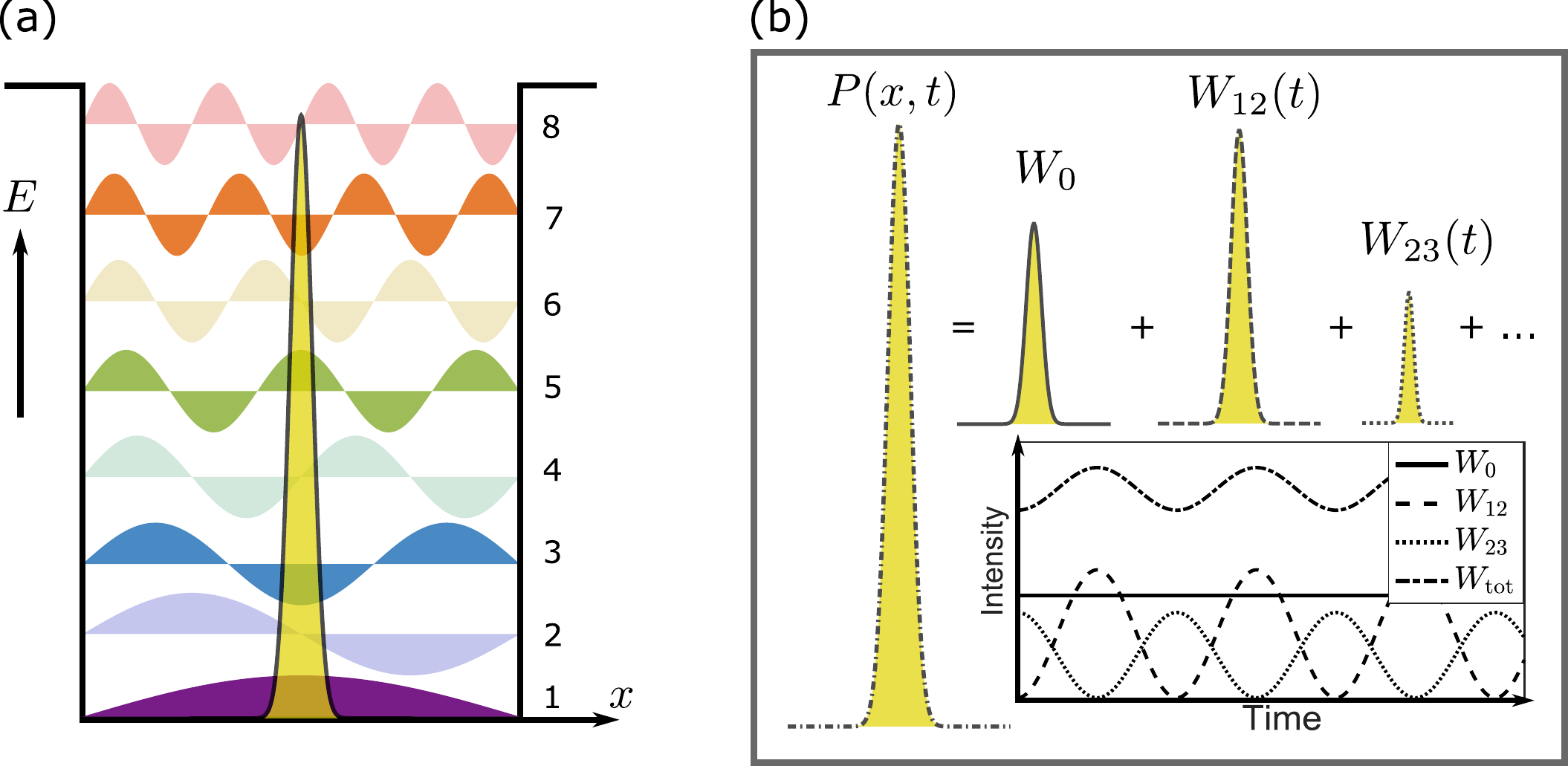}
\caption{(a) Energy-coordinate ($E$-$x$) schematic showing a polariton potential well with the first eight standing wave eigenstates $\varphi_m(x)$ and a delta shaped nonresonant pump $P(x,t)$ feeding particles into the system. The tight shape of the pump causes negligible overlap with odd states (low opacity) as opposed to even states (high opacity). (b) The pump is a superposition of many delta shaped pumps with different maximal intensities $W_{ij}$ oscillating with the frequency difference of the system eigenstates [see Eq.~\eqref{eq.pump}].}
\label{fig.scheme}
\end{figure}
\begin{equation}\label{eq:Schrodinger}
i\frac{\partial \Psi}{\partial t}=  \left[\hat{H}_0+ \alpha|\Psi|^2  + \frac{i}{2} \left( G P(x,t)(1 - g_I |\Psi|^2) - \Gamma \right)  \right] \Psi ,
\end{equation}
\begin{equation} \label{eq.pump}
P(x,t) = \delta(x) \left( W_0 + \sum_{i < j} W_{ij} \cos^2{\left( \frac{(\omega_i-\omega_j)t + \phi_{ij}}{2}\right)} \right),
\end{equation} 
where $G  = g_I - ig_R$ and $g_I,g_R>0$ are phenomenological parameters accounting for reservoir gain and blueshift respectively, $\Gamma$ is the inverse of polariton lifetime, $\alpha$ is the polariton-polariton interaction strength, and $\delta(x)$ is the Dirac-delta function. The oscillating nonresonant excitation is described by the weights $W_{ij}$ with phases $\phi_{ij}$, and $W_0$ corresponding to a static excitation source. Such a form of excitation can be arranged by driving a microcavity with a frequency comb of discrete frequencies~\cite{Gohle_PRL2007}. Previously, coherent frequency combs produced by the polariton system were predicted~\cite{Rayanov_PRL2015}, while driving of a polaritonic system at more than one frequency has been used to realize parametric amplifiers \cite{Savvidis_PRL2000}, two fluid-switches \cite{Giorgi_PRL2012}, or four-wave mixing spectroscopic techniques \cite{Kohnle_PRL2011}. We note that the main reason to assume a fast relaxing reservoir is to bring a single clear equation to the dynamics of the polariton system. The validity of Eq.~\eqref{eq:Schrodinger} is addressed in Sec.~S1 in the supplemental material (SM).

We write the order parameter in the basis of the bare eigenstates,
\begin{equation} \label{eq.ansatz}
\Psi(x,t)= \sum_{n=1}^N A_n(t) \varphi_n(x),
\end{equation}
where the coefficients $A_n(t)$ capture the dynamics of the condensate internal modes. Being a non-Hermitian problem the energies are complex where the net gain of the $n$-th mode, or its Lyapunov energy, is denoted $\lambda_n$.

By slowly increasing the values of the weights in time, the system will eventually condense when gain overtakes losses and the net gain becomes positive. This process relies on the order parameter finding the optimal solution through classical (thermal) fluctuations while in the uncondensed regime. The method is sometimes referred to as ground state \textit{approach from below}~\cite{Kalinin_2018ArXiv} where a solution with the lowest decay rate (highest $\lambda$) condenses ahead of others. Once found, bosonic stimulated scattering quickly populates the mode(s) to form a macroscopic condensate. 

We transform the macrocopic wavefunction to an appropriate basis and reduce our coordinate dependent complex Gross-Pitaevskii equation [Eq.~\eqref{eq:Schrodinger}] to a set of nonlinear equations describing the bosonic populations in each mode of the quantum well. The infinite quantum well eigenstates (standing wave basis) are written as
\begin{equation}
\varphi_n(x) = \sqrt{ \frac{2}{L} } \sin{ \left[n \pi \left( \frac{x}{L} + \frac{1}{2} \right) \right]}.
\end{equation}
The assumption of a spatially localized non-resonant pump means that odd parity states are not excited since they have no overlap with the pump. The problem thus reduces to only states of even parity and all indexing and summation is to be only taken over $n = 1,3,5,\dots$ from here on. Substituting Eq.~\eqref{eq.ansatz} into Eq.~\eqref{eq:Schrodinger}, integrating over the spatial coordinate and exploiting the orthogonality of the basis we get
\begin{align} \notag
 i\frac{\partial A_m}{\partial t} = &  i\frac{GP(t)}{L}  \left[ \sum_n p_{nm} A_n -  \frac{2 g_I}{L} \sum_{jkl} M_{mjkl}A_j^* A_k A_l \right]\\ \label{eq:Am}
    + & \left( \omega_m - i \frac{ \Gamma}{2} \right) A_m  +  \frac{\alpha}{2L} \sum_{jkl} T_{mjkl}A_j^* A_k A_l ,
\end{align}
where the delta-function pump gives a simple expression for the overlap elements $p_{nm} = \sgn{[\varphi_n(0) \varphi_m(0)]}$ and $M_{mjkl} = \sgn{[\varphi_m(0) \varphi_j(0) \varphi_k(0) \varphi_l(0)]}$. The nonlinear elements $T_{mjkl} \in \{3, 2, 1, -1, 0\}$ belonging to the polariton-polariton interaction term have a more complicated structure since the integration is not confined at origin. It is in principle possible to continue the development with a pump of different spatial shape with only the cost of calculating an extra set of overlap elements between the pump and the linear states. It is a good assumption that each mode possesses the same reservoir gain/saturation rate $g_I$ if the microcavity photons are sufficiently detuned from the exciton resonance (see Sec.~S2 in the SM).

Even in the absence of nonlinearity a general analytical solution method does not exist to Eq.~\eqref{eq:Am} due to the non-commutativity of the problem. Instead, we show an approximate method based on time-averaged equations of motion which can correctly predict solutions of largest net gain.

We assume that the optimal gain of the system belongs to a superposition of modes coupled together by the modulated excitation source analogous to active mode locking in lasers. Our ansatz is then written,
\begin{equation} \label{eq.ansatz2}
\Psi(x,t)= \sum_{n=1}^N [A_n + \delta_n(t)] \varphi_n(x) e^{-i \omega_n t}.
\end{equation}
Here we have assumed that the final state is characterized by a comb of energies $\hbar \omega_n$ whose average population $A_n$ experiences periodic fluctuations $\delta_n(t)$ whose contribution is zero in the time average limit. We point out that $A_n$ can depend on time, such as the transient process of going from an uncondensed state to condensed, but at a much slower rate than its characteristic frequencies, i.e. $\dot{A}_n \ll \omega_n A_n$. Furthermore, we work close to threshold in order to minimize the blueshift coming from polariton-polariton nonlinearities. 

Performing time averaging over fast oscillating terms around the mode average values $A_n$ and keeping only resonant terms we have
\begin{align} \notag
 i \frac{\partial A_m}{\partial \tau} 
 & = i \frac{G}{L} \bigg[ \left( W_{0} +  \sum_{n, m} \frac{W_{nm}}{2} \right)  \times \left(A_m -  \frac{2 g_I}{L} \sum_{\mathcal{G}_m} M_{mjkl}A_j^* A_k A_l  \right)  +  \sum_{\mathcal{T}_{m}}  \frac{W_{qk} }{4} p_{nm} A_{n_\pm} e^{\pm i\phi_{qk}} \\ \label{eq.AmAv}
&  -  \frac{ g_I}{2L} \sum_{\mathcal{K}_m} W_{nq}  M_{mjkl} A_{j_\pm}^* A_k A_l e^{\pm i  \phi_{nq}} \bigg] - i \frac{ \Gamma}{2} A_m  + \frac{\alpha}{2L} \sum_{\mathcal{G}_m} T_{mjkl}A_j^* A_k A_l .
\end{align}
Here, $\tau$ is a slow time variable such that $\partial_\tau W_{nq} \ll \omega_1 W_{nq}$. $\mathcal{T}_m, \mathcal{G}_m, \mathcal{K}_m$ are sets of indices which are solutions to the following diophantine equations, respectively,
\begin{align}\label{eq.triplet}
& n_\pm = \sqrt{m^2 \pm (q^2 - k^2)}, \\
& j = \sqrt{k^2 + l^2 - m^2}, \\
&j_\pm = \sqrt{k^2 + l^2 - m^2 \pm (n^2 - q^2)}.
\end{align}
The plus-minus notation necessarily arises to keep the coupling between modes symmetric. The first and second term in the RHS of Eq.~\eqref{eq.AmAv} (second curved bracket) correspond to gain and saturation due to the static nonresonant pump terms respectively. The third and fourth term correspond to gain and saturation coupling between different modes due to the oscillating pump terms respectively. The fifth term is the average dissipation equal for all modes. The sixth term represents polariton-polariton interactions.

Finally, we address the feasibility of the proposed scheme by considering the possible size of the all-to-all connected network. In modern GaAs samples~\cite{Sun_NatPhys2017} the polariton decay rate can be as low as $\hbar\Gamma = 6.5$ $\mu$eV and determines the minimum energy spacing between the system eigenmodes. The tight nonresonant beam profile only excites even modes of the system due to negligible overlap with odd modes. For this reason we are only interested in even modes of the system throughout the remainder of the paper. Let us then consider the infinite quantum well as depicted in Fig.~\ref{fig.scheme}a. A spacing $E_3-E_1 = 6.5$ $\mu$eV between the ground state and second excited state can be achieved for a polariton mass $m^* = 10^{-4} \times m_0$, where $m_0$ is the free electron mass, and the well of width $L = 70$ $\mu$m. Operating over a 5 meV bandwidth (less than the GaAs exciton binding energy) the maximum possible mode quantum number is then $n \approx 80$ giving possible control over 40 even modes through the nonresonant excitation. Even though this may not seem like many, we note that previous examples of polariton networks have been restricted to nearest neighbour type coupling. It is generally expected that nearest neighbour coupled systems need to be orders of magnitude larger in size to represent the same complexity as all-to-all coupled networks (for example, $40$ all-to-all coupled nodes would need $\sim10^4$ nodes to be represented in a nearest neighbour graph~\cite{Cueves_Sci2016}). Higher numbers of modes could be feasible in two-dimensional geometries or other material systems, e.g., carbon nanotube based polaritons have shown record Rabi splitting exceeding $300$~meV allowing to operate over a wider bandwidth~\cite{Gao_NatPho2018}.

\section{Results}
In this section we present results on the dynamics of the polariton condensate by slowly increasing the excitation intensity and populating the state of highest gain. We then compare the results with optimal states predicted by the time averaged theory. We note that below condensation threshold the dynamics of the wavefunction are determined by a weak stochastic field (not shown in equations) in congruity with the truncated Wigner approach~\cite{Wouters_2009PRB}.
\begin{figure}[t!]
\centering
\includegraphics[width=0.7\linewidth]{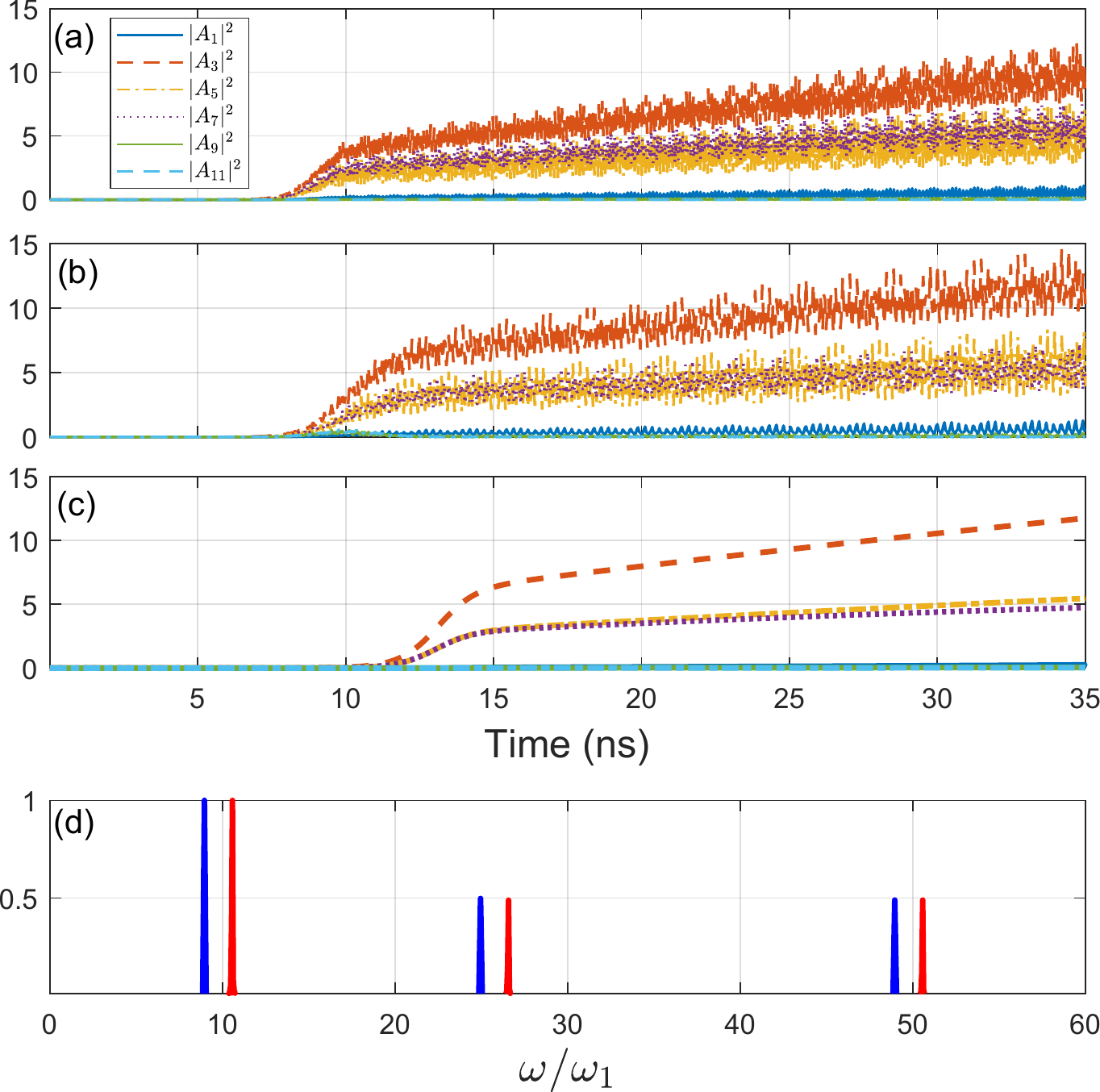}
\caption{Time integrated results from Eqs.~\eqref{eq:Schrodinger},\eqref{eq:Am}, and \eqref{eq.AmAv} in panels (a),(b), and (c) respectively. Weights are increased slowly in time to their mark values $W_0 = 0.4$ $\mu$m$^{-1}$ and $W_{35} = W_{37} = 3W_0$. Pump phases were set to $\phi_{35} = \pi$ and $\phi_{37} = 0$. The oscillating condensed state is characterized by a hierarchy predicted by Eq.~\eqref{eq.AmAvState} as the optimal state of the time averaged couplings between different modes. (d) Comparison of the normalized spectrum from the final state in panel (b) [red], and Eq.~\ref{eq.AmAvState} [blue]. In our calculations there is a small blueshift due to polariton-polariton interactions, which remains limited as we operate only slightly above threshold.}
\label{fig.AmAv}
\end{figure}

Let us consider the $N=6$ mode system and investigate the example of a nonresonant pump slowly increased in time to the mark values $W_0 = 0.4$ $\mu$m$^{-1}$ and $W_{35} = W_{37} = 3W_0$, with phases $\phi_{35} = \pi$ and $\phi_{37} = 0$. The linear time averaged coupling then looks like explicitly as
\begin{equation} \label{eq.AmAvMat}
\mathbf{H} =  \frac{G}{4L} \begin{pmatrix}
W & 0 		& 0 			& 0 			& 0 			& 0 \\
0 & W 		& W_{35} 	& W_{37} 	& 0 			& 0 \\
0 & W_{35} &  W 		& 0 			& 0 			& 0 \\
0 & W_{37} & 0 			& W 		& 0 			& 0 \\
0 & 0 		& 0 			& 0 			& W 		& -W_{37} \\
0 & 0 		& 0 			& 0 			& -W_{37} & W 
\end{pmatrix},
\end{equation}
where $W = 4W_0 + 2(W_{35}+W_{37} - \Gamma L/G)$. It becomes now clear why the pump [Eq.~\eqref{eq.pump}] was chosen with such time dependence. Weights $W_{qk}$ now act as couplings between modes $A_q$ and $A_k$. Using similar reasoning with extra nonresonant pump then allows for the realization of an all-to-all coupled dissipative bosonic network. Solving the eigenvalue problem of this matrix gives a maximal $\lambda$ for the following eigenvector,
\begin{equation} \label{eq.AmAvState}
\mathbf{A} =\begin{pmatrix} 
A_1 \\
A_3 \\
A_5 \\
A_7 \\
A_9 \\
A_{11}
\end{pmatrix} =
 \frac{1}{2}\begin{pmatrix} 
0 \\
\sqrt{2} \\
1 \\
1 \\
0 \\
0
\end{pmatrix}
\end{equation}
Performing numerical integration of Eqs.~\eqref{eq:Schrodinger},\eqref{eq:Am},\eqref{eq.AmAv} for the chosen weights and keeping otherwise previous parameters unchanged~\cite{parameters} we plot the results in Fig.~\ref{fig.AmAv}(a-c) respectively. As expected the condensed state is composed of dominant populations in $A_3$, $A_5$ and $A_7$ which have the highest gain on average with the oscillating pump. The slight mismatch between panel (a) and (b) comes from the fact that a finite width pump (FWHM $\sim 1.4$~$\mu$m) is used in Eq.~\eqref{eq:Schrodinger} whereas delta peak pump is assumed in Eqs.~\eqref{eq:Am}, \eqref{eq.AmAv}. The hierarchy of the bosonic populations is predicted correctly by Eq.~\eqref{eq.AmAvState} showing the abilities of the system to find the optimal state in the time average. We additionally show in Fig.~\ref{fig.AmAv}d comparison between the spectrum of the final state in Fig.~\ref{fig.AmAv}b (red lines) and Eq.~\ref{eq.AmAvState} (blue lines) respectively. Results from all stages in the theory show good agreement with the predicted optimal state [Eq.~\eqref{eq.AmAvState}] underlining that the system works as an optimizer for the coupled equations of motion where one can deterministically create specific bosonic distributions by an appropriate choice of weights.
\begin{figure}[t!]
\centering
\includegraphics[width=0.7\linewidth]{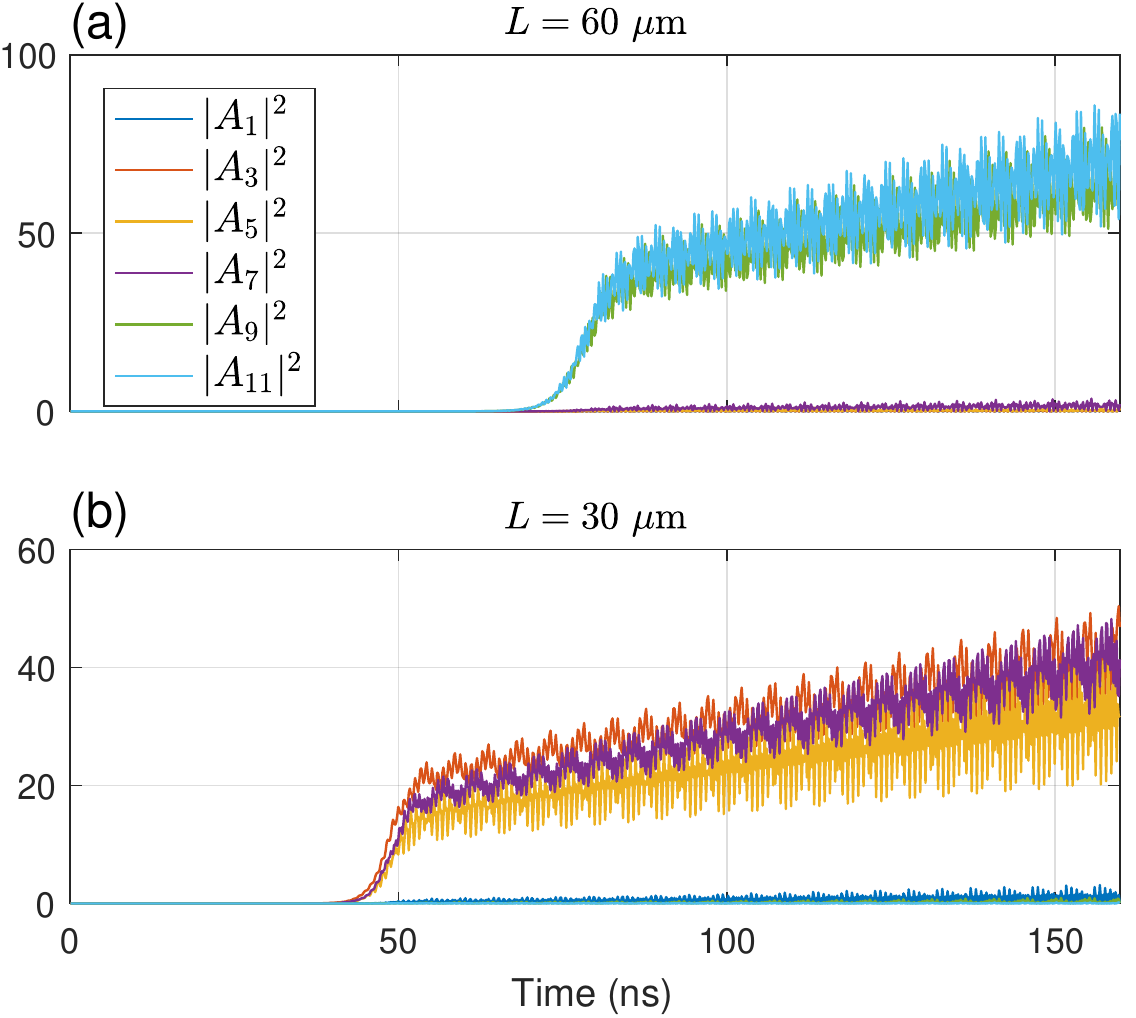}
\caption{A repeat of the simulation presented in Fig.~\ref{fig.AmAv}b. Here, we set $g_R = 4 \alpha$ in accordance with the Hartree-Fock theory and $g_I = 0.008$ ps$^{-1}$ $\mu$m in order to produce a condensation blueshift $\sim 100$ $\mu$eV. In panel (a) the trap width is $L=60$ $\mu$m and in (b) $L = 30$ $\mu$m. For the former the system finds a state corresponding to the second biggest eigenvalue of Eq.~\eqref{eq.AmAvMat} whereas the latter shows a resemblance to the optimal state of the time average, demonstrating that when the energies $\omega_n$ are small then perturbation from the pump induced potential ($g_R P(t)/L$) affects the outcome.}
\label{fig.AmAv2}
\end{figure}

We point out that in the time average the reservoir blueshift $g_R$ does not affect the optimal solution coming from the linear terms in Eq.~\eqref{eq.AmAv}. However, in Eq.~\eqref{eq:Am} the presence of a pump induced potential $g_RP(t)/L$ perturbs the trap dispersion $\omega_n$ and consequently causes discrepancy between the output solution from Eq. \eqref{eq:Am} and the optimal time average solution from Eq. \eqref{eq.AmAv}. As an example, in Fig.~\ref{fig.AmAv2} we repeat the simulation presented in Fig.~\ref{fig.AmAv}b, but now choose $g_R = 4 \alpha$ in accordance with the Hartree-Fock theory~\cite{Wouters_PRL2007} and $g_I = 0.008$ ps$^{-1}$ $\mu$m. This choice of parameters produces a condensation blueshift of $\sim 100$ $\mu$eV and a strong pump induced potential of $\sim 200$ $\mu$eV similar to experiments in GaAs systems~\cite{Kasprzak_Nature2006} and previous theoretical works~\cite{Ostrovskaya_2013PRL}. In Fig.~\ref{fig.AmAv2}a the resulting condensed state is characterized by approximately equal populations in $A_9$ and $A_{11}$ mode corresponding to the second biggest eigenvalue of Eq.~\eqref{eq.AmAvMat}.  By increasing the system energies (e.g. decreasing trap width $L$) and minimizing the perturbing effects of the pump induced potential one retrieves the optimal solution as is shown in Fig.~\ref{fig.AmAv2}b. Exact calculations on the perturbed dispersion $\omega_n'$ is beyond the scope of this paper and we focus on the case where this perturbation is small. 

\section{Benchmarking}
The ability of the system to produce an output in agreement with the optimal state coming from the linear couplings of Eq. \eqref{eq.AmAv} is influenced by the pump induced potential $g_R$ and nonlinear effects upon condensation. This influence can be characterized by benchmarking both Eq.~\eqref{eq:Am} and Eq.~\eqref{eq.AmAv} considering different sparsity of couplings. We will set $W_0 = 0$ since it creates equal gain for all modes and is therefore not important. Other parameters are given in Ref.~\cite{parameters}. We limit ourselves to a system of first six even modes $\{A_1,A_3,A_5,A_7,A_9,A_{11}\}$ which can be driven by 15 distinct weights. Each numerical trial uses a random set of weights and random initial conditions. The weights are picked from a uniform random distribution $\in [0,1]$, and are normalized consequently by their sum. This ensures that the net intensity of the nonresonant excitation is fixed in every random trial.
\begin{figure*}[t!]
\centering
\includegraphics[width=1\linewidth]{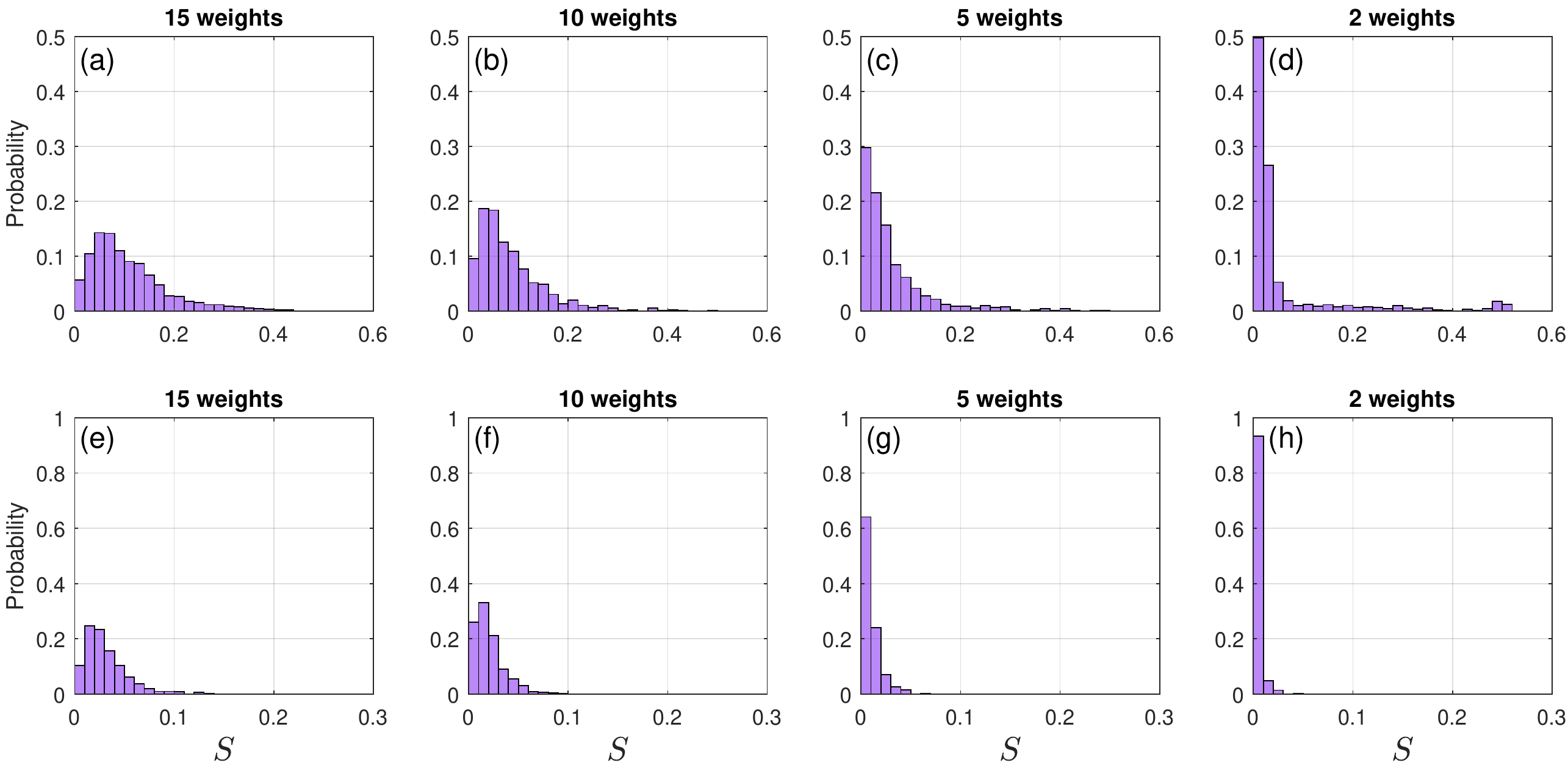}
\caption{(a-d) Benchmarking of Eq.~\eqref{eq:Am} and (e-h) Eq.~\eqref{eq.AmAv} showing the probability distribution for the energy deviation from the optimal (maximal eigenvalue) solution $S(\mathbf{A})$. We used 1000 random trials in each panel and considered different sparsity of the problem.}
\label{fig.BM}
\end{figure*}

As in Fig.~\ref{fig.AmAv} we slowly raise the value of the weights to their mark values. At the end of each trial we measure the success of the equations in producing the optimal state from the linear couplings given by the weights. The linear coupling matrix $\mathbf{J}$  to be benchmarked comes from the third term in Eq.~\eqref{eq.AmAv},
\begin{equation}
J_{nm} =  \sum_{\mathcal{T}_{m}}  W_{qk}  p_{nm} A_{n_\pm} e^{\pm i\phi_{qk}},
\end{equation}
which is analogous to $\mathbf{H}$ in Eq.~\eqref{eq.AmAvMat} without diagonal elements and a multiplication factor (which does not affect the hierarchy of the eigenvalues). A Lyupunov energy $\lambda$ is associated with $\mathbf{J}$ for some given vector $\mathbf{A}$,
\begin{equation}
\lambda_{\mathbf{A}} =  \frac{(\mathbf{A}, \mathbf{J} \mathbf{A})}{(\mathbf{A},\mathbf{A})},
\end{equation}
where the brackets denote inner product. The matrix $\mathbf{J}$ is Hermitian with maximum and minimum real eigenvalue denoted $\lambda_\text{max}$ and $\lambda_\text{min}$, respectively, which are found using the QR algorithm. We then define the normalized distance from the maximum eigenvalue as
\begin{equation}
S(\mathbf{A}) =\frac{\lambda_\text{max} - \lambda_\mathbf{A}}{\lambda_\text{max} - \lambda_\text{min}}.
\end{equation}
Results are presented in Fig.~\ref{fig.BM}(a-b) and Fig.~\ref{fig.BM}(e-f) for Eq.~\eqref{eq:Am} and Eq.~\eqref{eq.AmAv} respectively for a different number of weights in the system. We note that generally the fewer the number of weights the more sparse the problem becomes and the chances of finding the optimal solution increase. The specific benchmarking scenario where $g_R$ and/or $\alpha$ are zero is addressed in Sec.~S3 in the SM. We point out the different scales on the vertical axes between the upper and the lower panels. It is not surprising to see a greater success in the time averaged model since it already assumes that the system will find a solution of the form given by Eq.~\eqref{eq.ansatz2}. The effects of various errors in the nonresonant excitation is addressed in Sec.~S4 in the SM. Overall we find good performance of the system where the average deviation from all panels Fig.~\ref{fig.BM}[a-d] is $\langle S \rangle \approx 8 \%$.

\section{Conclusions}
We have studied a method of creating a network of all-to-all coupled bosonic modes in a trapped condensate of exciton-polaritons. An optimal solution to the network corresponds to a lasing state of lowest decay (optimal gain) found by an \textit{approach from below} method. This method relies on slowly activating an external excitation source which allows the condensate to form in a solution of optimal gain similar to active mode locking.

We show that the couplings can be realized using a nonresonant excitation source with oscillating intensity at resonance with the trap energy level spacing. The couplings between bosonic modes are then directly tunable via the excitation method. This allows one to create optimal gain conditions for a state characterized by a distribution of polaritons in selected trap modes. We showed that the system dynamically simulates the competition process between modes, and solves a max-eigenvalue problem for dense matrices $\mathbf{J}$.

The outlook towards future investigations can include biased problems where an additional set of coherent beams create, on average, a nonhomogeneous problem for the condensate equation of motion. Recently, phase modulated optical resonators were theoretically shown to produce dynamics analogous to the Haldane model~\cite{Yuan_PRB2018} which raises the question whether modulated polariton traps are suitable for such synthesized lattices. Also, investigation into implementing constrained problems where the system settles for a minimum in a limited state space would be advantageous in the context of quadratic programming problems,~\cite{KOZLOV_CMMP1980} which are related to optimization of NP-hard complex systems~\cite{Pardalos_JGO1991}. Such constraints can possibly be introduced by additional nonlinear terms to the equations of motion derived from an appropriate Hamiltonian density and will be the subject of our future works.

\begin{acknowledgement}
H.S. acknowledges support by the Research Fund of the University of Iceland, The Icelandic Research Fund, Grant No. 163082-051. K.D. acknowledges support from 5-100 program of the government of Russian Federation. T.L. was supported by the Singaporean MOE grant No. 2017-T2-1-001.
\end{acknowledgement}

\setcounter{secnumdepth}{1} 
\renewcommand{\thefigure}{\textbf{S\arabic{figure}}}
\renewcommand{\thesection}{S\arabic{section}}
\renewcommand{\theequation}{S\arabic{equation}}

\begin{suppinfo}

\section{Quasi-stationary reservoir approximation}
The equations of motion for the polariton field $\Psi$ and active exciton reservoir $n_R$ are written as~\cite{Wouters_PRL2007}:
\begin{align}
i\frac{\partial \Psi}{\partial t} & =  \left[\hat{H}_0+ \alpha|\Psi|^2  + \frac{i}{2} \left( G' n_R  - \Gamma \right)  \right] \Psi ,\\
\frac{\partial n_R}{\partial t} & = -(\Gamma_R + g_I' |\Psi|^2) n_R + P(x,t).
\end{align}
Here $G' = g_I' - i g_R'$ and $g_I',g_R'>0$ are phenomenological parameters depicting polariton scattering rate from the reservoir into the condensate and reservoir interactions respectively, $\Gamma$ and $\Gamma_R$ are the polariton and reservoir inverse lifetimes, respectively, and $\alpha$ is the polariton-polariton interaction strength.

Assuming that the reservoir relaxes much faster than the condensate and follows the excitation intensity, we can apply a quasi-stationary approximation $\partial_t n_R = 0$ which tells us that at each moment in time the reservoir follows the pump intensity. This is valid when $\Gamma_R$ is much larger than the frequencies characterizing the pump. Additionally, staying close to the condensation threshold (low condensate intensity $|\Psi|^2$) a Taylor expansion of the stationary reservoir gives
\begin{equation} \label{eq.nR0}
n_R^{(0)} = \frac{P(x,t)}{\Gamma_R + g_I' |\Psi|^2} \approx \frac{P(x,t)}{\Gamma_R} \left(1 - \frac{g_I' |\Psi|^2}{\Gamma_R} \right).
\end{equation}
In Fig.~\ref{fig.nR} we show results on the validity of Eq.~\eqref{eq.nR0} neglecting spatial degrees of freedom. We have chosen $\Gamma = 0.01$ ps$^{-1}$, $\Gamma_R = 100\Gamma$, $g_I' = \alpha = 0.001$ ps$^{-1}$ $\mu$m, $g_R = 10g_I$. Here we write the pump as
\begin{equation}
P(t) = W_0 + W_{13} \cos^2{ \left( \frac{\omega_1 - \omega_3}{2} t  \right)} +  W_{15} \cos^2{ \left( \frac{\omega_1 - \omega_5}{2} t  \right)},
\end{equation}
where $\hbar \omega_1 = 6.5$ $\mu$eV is the ground state energy of the infinite potential well. A maximum period is then $T = 2\pi/(\omega_3 - \omega_1) \approx 79.6$ ps. Fixing $W_{13} = 1$ ps$^{-1}$ $\mu$m$^{-1}$ and $W_{15} = 2W_{13}$ we find that threshold takes place around $ W_0 = W_\text{th} \approx 8.55$ ps$^{-1}$ $\ium$. Below threshold ($W_0 = 0.93W_\text{th}$) only the reservoir is active in the sytem [Fig.~\ref{fig.nR}a]. Slightly above threshold ($W_0  = 1.05W_\text{th}$) the condensate forms, with Eq.~\eqref{eq.nR0} remaining valid [Fig.~\ref{fig.nR}b]. Further above threshold ($W_0 = 1.17W_\text{th}$) a deviation between the true reservoir $n_R$ and the approximate reservoir $n_R^{(0)}$ becomes apparent [Fig.~\ref{fig.nR}c]. 
\begin{figure}[h!]
\includegraphics[width = 0.8\linewidth]{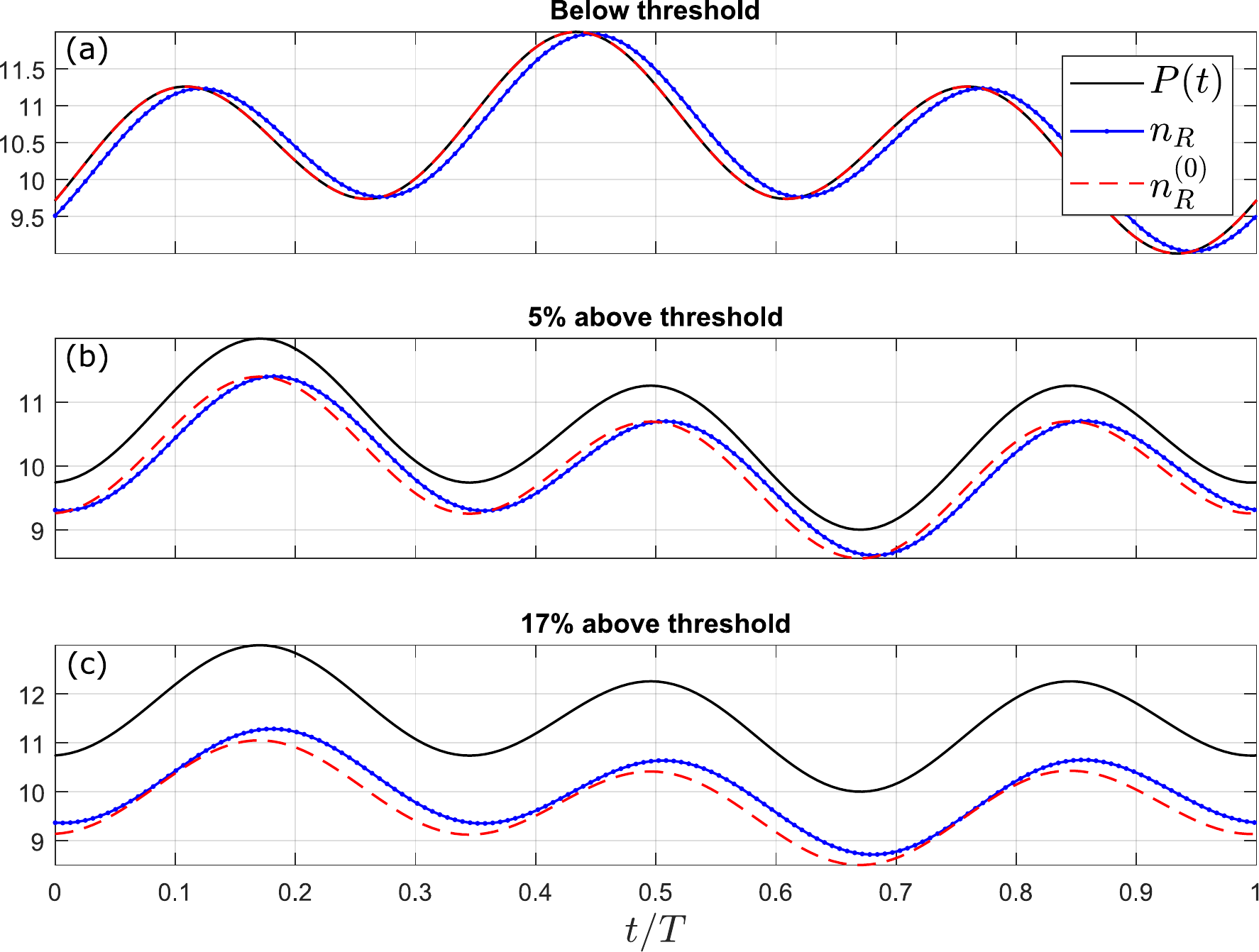}
\caption{Dynamics of the true reservoir $n_R$ against the quasi-stationary approximated reservoir $n_R^{(0)}$ for an oscillating nonresonant pump. (a) Below condensate threshold, (b) 5\% above threshold, and (c) 17\% above threshold. Results show that when $\Gamma_R$ is chosen sufficiently large and one stays close to threshold the quasi-stationary approximation is valid ($n_R \approx n_R^{(0)}$).}
\label{fig.nR}
\end{figure}

\section{Mode dependent gain rates}
We investigate the effects of mode dependent gain/saturation rates. The origin of different rates from the reservoir stems from varying Hopfield coefficients of the polariton state. For larger modes the excitonic fraction becomes more dominant and consequently experiences a higher saturation rate from the reservoir. If the cavity photon mode is detuned from the exciton reservoir then this change in Hopfield fractions can be minimal as is shown in Fig.~\ref{fig.diffSat}a. In the figure we show the first 8 even eigenmodes of the infinite quantum well in reciprocal space (colormap) and the lower polariton branch (black dashed line) in a trap-free system, and the exciton (blue dot-dashed line) and photon (red dotted line) Hopfield coefficients. Here we have chosen a Rabi splitting of 4 meV and negative detuning of -10 meV, and a cavity photon mass $m^* = 5m_0  \times 10^{-5}$ where $m_0$ is free electron mass. 

As a first approximation we take the change in the saturation rates as linear in mode energy where Eq.~\eqref{eq:Am} in the main text is now written
\begin{align} \notag
 i\frac{\partial A_m}{\partial t} = &  i\frac{P(t)}{L}  \left[ \sum_n (g_I w_{nm} - ig_R p_{nm}) A_n -  \frac{2 g_I}{L} \sum_{jkl} (g_I N_{mjkl} - i g_R M_{mjkl})A_j^* A_k A_l \right]\\ \label{eq:AmGR}
    + & \left( \omega_m - i \frac{ \Gamma}{2} \right) A_m  +  \frac{\alpha}{2L} \sum_{jkl} T_{mjkl}A_j^* A_k A_l.
\end{align}
Here,  $M_{mjkl} = \sgn{[\varphi_m(0) \varphi_j(0) \varphi_k(0) \varphi_l(0)]}$ and $p_{nm} =\sgn{[\varphi_n(0) \varphi_m(0)]}$ as usual but we write the new elements $N_{njkl}$ and $w_{nm}$ as
\begin{align} \label{eq.NLin}
N_{mjkl} &= \sgn{[\varphi_m(0) \varphi_j(0) \varphi_k(0) \varphi_l(0)]} \times [1 + \epsilon (m^2 + j^2 + k^2+ l^2)],\\
w_{nm} &= \sgn{[\varphi_n(0) \varphi_m(0) ]} \times [1 + \epsilon (n^2 + m^2 )].
\end{align}
Here, $\epsilon$ is a tunable parameter to investigate the effect of modes experiencing gain and saturation at different rates. Results are presented in Fig.~\ref{fig.diffSat} where we show the dynamics analogous to Fig.~\ref{fig.AmAv}b in the main text but using Eq.~\eqref{eq:AmGR} with $\epsilon = 0.001$ (b), and $\epsilon = 0.002$ (c). Results show when $\epsilon$ is sufficiently small the original optimal solution is retrieved [Fig.~\ref{fig.diffSat}b]. When $\epsilon$ is increased a different state becomes optimal due to the coupling matrix, Eq.~\eqref{eq.AmAvMat} in main manuscript, being altered as predicted by the time average theory.

We finally point out that in the case where different modes experience different lifetimes on average then we replace $\Gamma \to \Gamma_n$. If $\Gamma_n$ increases fast with mode number then the system starts favoring populations only in low energy modes and it becomes increasingly difficult to tailor states with arbitrary time average bosonic distributions.
\begin{figure}[h!]
\includegraphics[width = 1\linewidth]{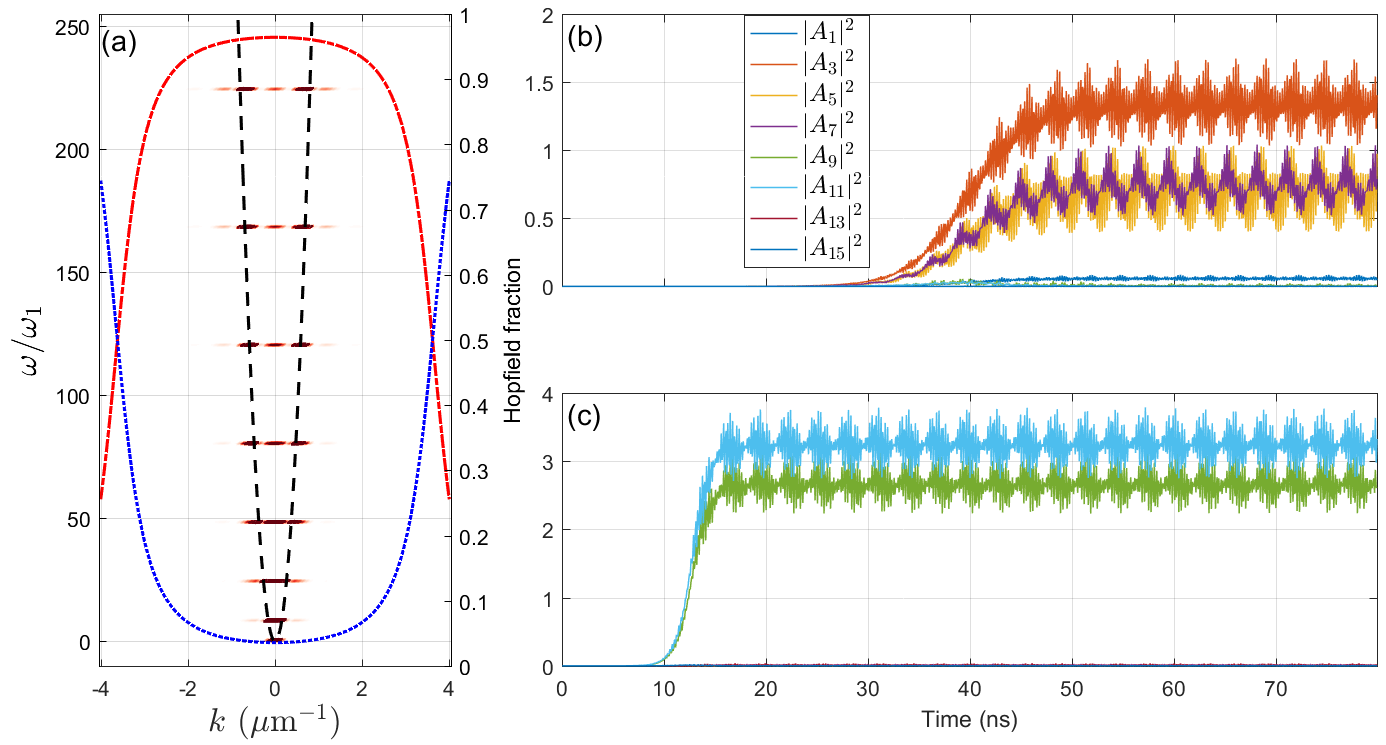}
\caption{(a) First 8 even parity eigenmodes of the infinite quantum well of width $L = 60$ $\mu$m in reciprocal space (colormap) with the lower polariton branch (black dashed line) in a trap free system, and the exciton (blue dot-dashed line) and photon (red dotted line) Hopfield coefficients. (b) Repeat of simulations from Fig.~\ref{fig.AmAv}b in main manuscript using Eq.~\eqref{eq:AmGR} with $\epsilon = 0.001$ and (c) $\epsilon = 0.002$. In both cases the time average theory correctly predicts the optimal state which are different in each panel because of the gain rates biasing the weights towards a solution occupying higher modes.}
\label{fig.diffSat}
\end{figure}

\section{Benchmarking with $\alpha$ and/or $g_R$ zero}
In Fig.~\ref{fig.BM2} we investigate the success probability of Eq.~\eqref{eq:Am} in the main text when $\alpha$ and/or $g_R$ are neglected. For the given set of parameters\cite{parameters} the results show greatly enhanced performance when reservoir blueshift is absent (Fig.~\ref{fig.BM2}[b,d]), but when only polariton-polariton interactions are absent (Fig.~\ref{fig.BM2}c) they are almost unchanged. The perturbation coming from the pump induced potential is therefore the dominant cause of deviation between the output state and the expected optimal state. This is in favor of systems where the reservoir gain rate $g_I$ is a dominant parameter. In organic microcavities the gain rate has been reported much higher then in inorganic GaAs microcavities (see e.g. review by Sanvitto and K\'ena-Cohen~\cite{Sanvitto_NatMat2016}) and could therefore be ideal candidates for creating the dissipative polariton networks suggested here.
\begin{figure}[t!]
\centering
\includegraphics[width=0.7\linewidth]{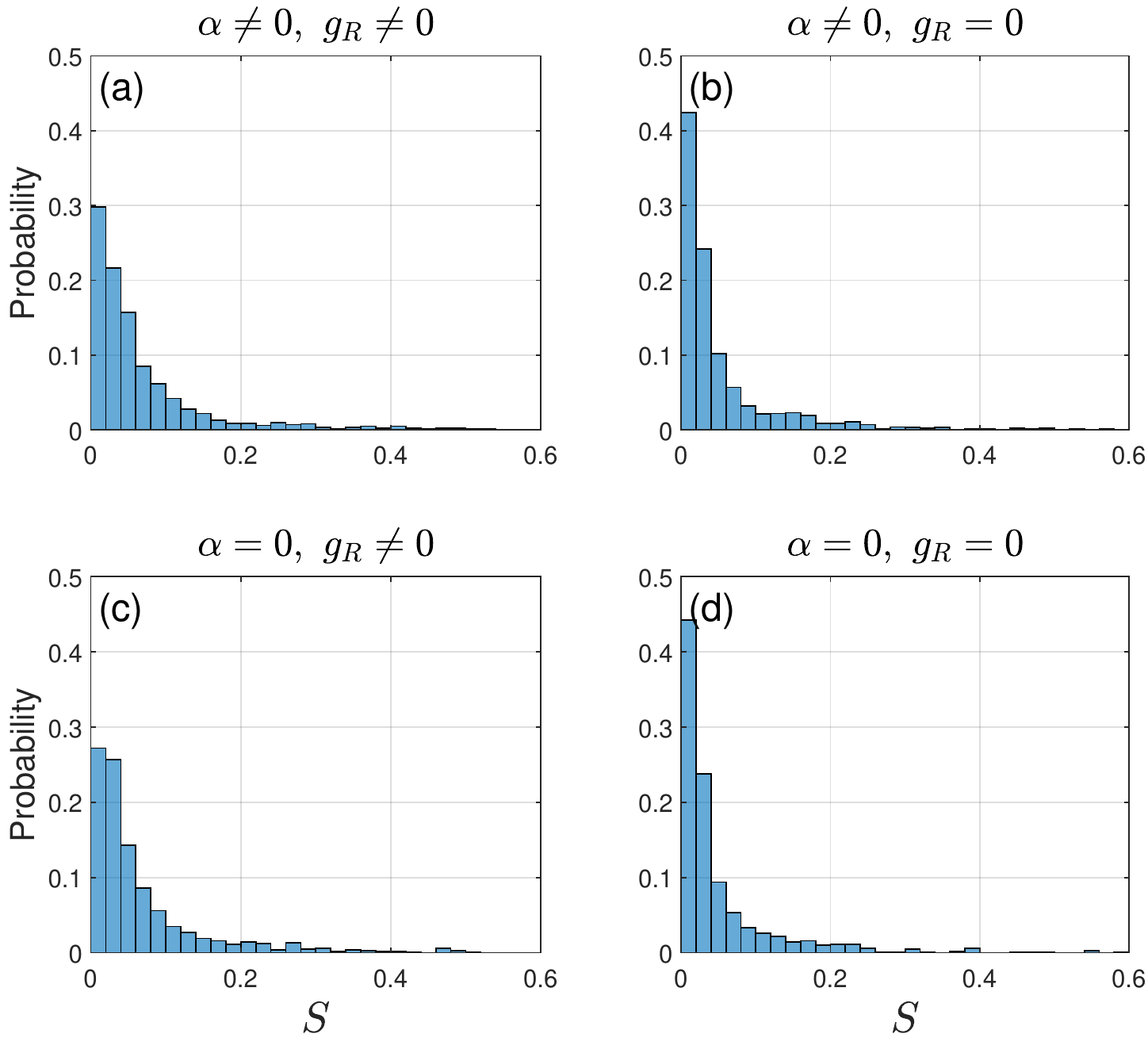}
\caption{Benchmarking of Eq.~\eqref{eq:Am} in main text for 5 weights with different configuration of $g_R$ and $\alpha$ given in each panel.}
\label{fig.BM2}
\end{figure}

\section{Modulational errors}
In this section we investigate the effects of errors in the modulation element (nonresonant excitation beam) on the performance of the system in finding the expected state of optimal gain. First, error in the amplitudes of various frequencies (weight error) can be written
\begin{equation}\label{eq.err1}
W_{ij}' = W_{ij} \times (1 + \theta_{ij}^W),
\end{equation}
where $\theta_{ij}^{W}$ is a uniformly distributed random variable on the interval $[-\Delta, \Delta]$ with $\Delta < 1$.

Second, error in the frequencies of the weights is introduced as,
\begin{equation}\label{eq.err2}
\omega_i' = \omega_i + \omega_1 \theta_i^{F}(\sigma),
\end{equation}
where $\theta_i^{F}(\sigma)$ is a normal distributed random variable with standard deviation $\sigma$. 

Third, error when unwanted weights are generated,
\begin{equation}\label{eq.err3}
\sum_{ij} [(1-D)W_{ij} + D \mathcal{W}_{ij}] = \text{const}
\end{equation}
where $\mathcal{W}_{ij}$ are unwanted weights corresponding to frequency peaks with random phase, random frequency within the bandwidth of the problem, and random amplitude. The coefficient $D$ corresponds to the fraction of input power going into the unwanted weights.

Results are shown in Fig.~\ref{fig.BM3} for various strength of error. In Fig.~\ref{fig.BM3}[a-b] we have $\Delta = 0.1, \ 0.3, \ 0.5$ respectively. In [d-e] we have $\sigma = 0.001, \ 0.003, \ 0.005$ respectively. In [g-i] we have $D = 0.1, \ 0.3, \ 0.5$ respectively. The system shows low sensitivity towards changes in the weights [a-b] and the presence of unwanted weights [g-i] indicated by the low drop in success. Changes in the frequencies show a higher sensitivity where fluctuations around $\sim 1 \%$ of the ground state energy ($\omega_1$) can dramatically decrease the performance [d-f]. In our setup of $L= 60$ $\mu$m and typical polariton mass, this would correspond to a $\sim 1/2$ GHz error in frequencies.

\begin{figure}[t!]
\centering
\includegraphics[width=1\linewidth]{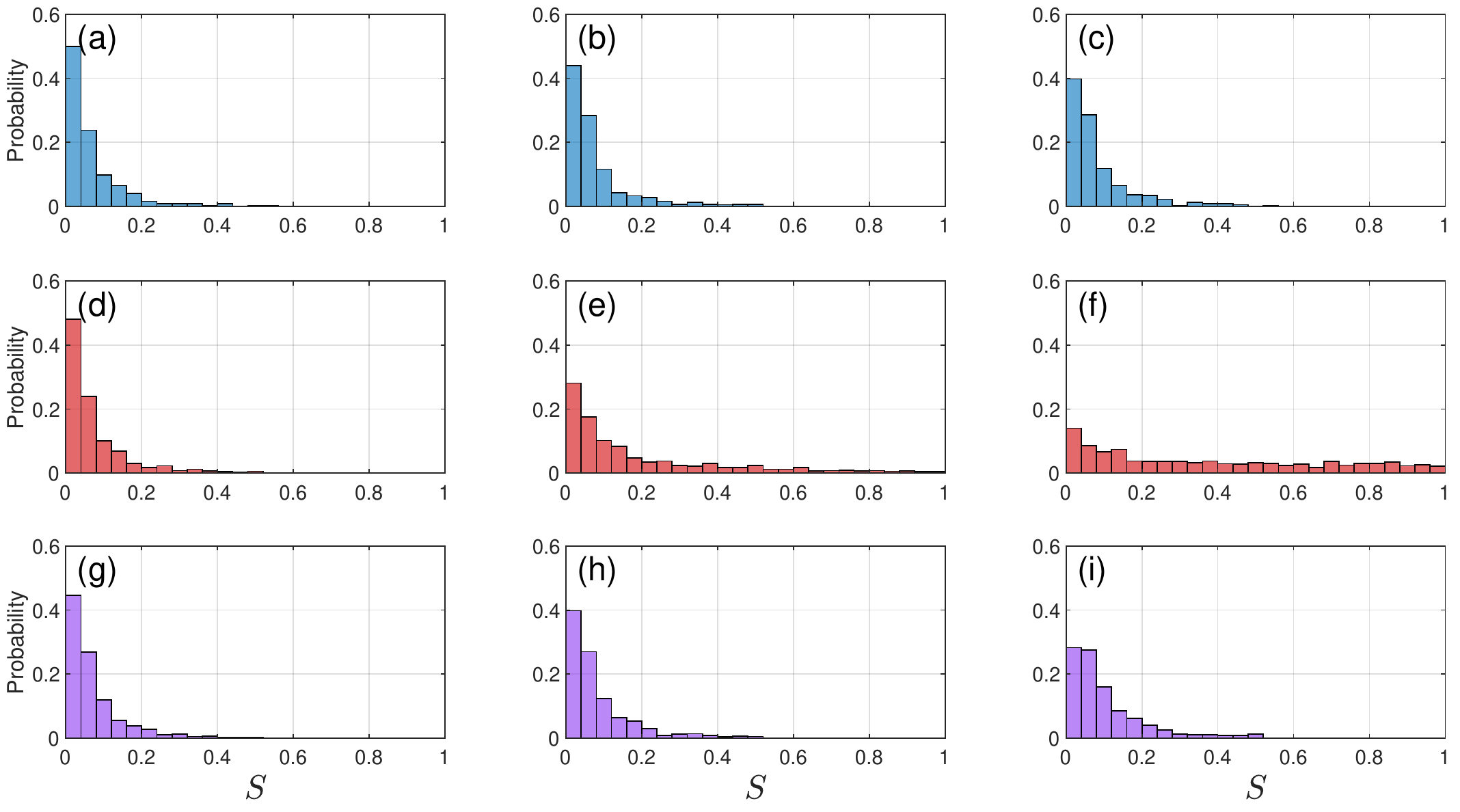}
\caption{Benchmarking of Eq.~\eqref{eq:Am} in main text for 5 weights. Errors are introduced using Eqs.~\eqref{eq.err1}-\eqref{eq.err3} in top, middle and bottom row of panels respectively. (a-c) $\Delta = 0.1, \ 0.3, \ 0.5$, (d-f) $\sigma = 0.001, \ 0.003, \ 0.005$, (g-i) $D = 0.1, \ 0.3, \ 0.5$.}
\label{fig.BM3}
\end{figure}

\end{suppinfo}

\bibliography{Bibliography}

\begin{mcitethebibliography}{52}
\bibitem[par()]{parameters}
Parameters: $\Gamma = 0.01$ ps$^{-1}$, $m^* = 5 \times 10^{-5} m_0$,
  $\hbar\alpha = 2.4$ $\mu$eV $\mu$m, $g_I = 0.2$ ps$^{-1}$ $\mu$m, $g_R = 0.1$
  ps$^{-1}$ $\mu$m. The choice of $g_R$ was taken to produce a pump induced
  potential of maximum blueshift $\sim 100$ $\mu$eV. The parameter $g_I$ was
  chosen to produce small nonlinearities at condensation.\relax
\mciteBstWouldAddEndPunctfalse
\mciteSetBstMidEndSepPunct{\mcitedefaultmidpunct}
{}{\mcitedefaultseppunct}\relax
\EndOfBibitem
\end{mcitethebibliography}

\end{document}